# Fungal bioremediation of diuron-contaminated waters: evaluation of its degradation and the effect of amendable factors on its removal in a trickle-bed reactor under non-sterile conditions

Kaidi Hu[1]; Josefina Torán[1]; Ester López-García[2]; Maria Vittoria Barbieri[2]; Cristina Postigo[2]; Miren López de Alda[2]; Gloria Caminal[3]; Montserrat Sarra[1*]; Paqui Blánquez[1]

[1]*Departament d'Enginyeria Química, Biològica i Ambiental, Escola d'Enginyeria, Universitat Autònoma de Barcelona, 08193 Bellaterra, Barcelona, Spain*

[2]*Water and Soil Quality Research Group, Department of Environmental Chemistry, Institute of Environmental Assessment and Water Research (IDAEA), Spanish Council for Scientific Research (CSIC), Jordi Girona 18-26, 08034 Barcelona, Spain*

[3]*Institut de Química Avançada de Catalunya (IQAC), CSIC. Jordi Girona 18-26, 08034 Barcelona, Spain*

***Corresponding author:***

Departament d'Enginyeria Química, Biològica i Ambiental, Escola d'Enginyeria, Universitat Autònoma de Barcelona, 08193 Bellaterra, Barcelona, Spain

Tel: +34 935812789, montserrat.sarra@uab.cat

**Abstract**

The occurrence of the extensively used herbicide diuron in the environment poses a severe threat to the ecosystem and human health. Four different ligninolytic fungi were studied as biodegradation candidates for the removal of diuron. Among them, T. versicolor was the most effective species, degrading rapidly not only diuron (83%) but also the major metabolite 3,4-dichloroaniline (100%), after 7-day incubation. During diuron degradation, five transformation products (TPs) were found to be formed and the structures for three of them are tentatively proposed. According to the identified TPs, a hydroxylated intermediate 3-(3,4-dichlorophenyl)-1-hydroxymethyl-1-methylurea (DCPHMU) was further metabolized into the N-dealkylated compounds 3-(3,4-dichlorophenyl)-1-methylurea (DCPMU) and 3,4-dichlorophenylurea (DCPU). The discovery of DCPHMU suggests a relevant role of hydroxylation for subsequent N-demethylation, helping to better understand the main reaction mechanisms of diuron detoxification. Experiments also evidenced that degradation reactions may occur intracellularly and be catalyzed by the cytochrome P450 system. A response surface method, established by central composite design, assisted in evaluating the effect of operational variables in a trickle-bed bioreactor immobilized with T. versicolor on diuron removal. The best performance was obtained at low recycling ratios and influent flow rates. Furthermore, results indicate that the contact time between the contaminant and immobilized fungi plays a crucial role in diuron removal. This study represents a pioneering step forward amid techniques for bioremediation of pesticides-contaminated waters using fungal reactors at a real scale.

## 1. Introduction

Over the last few decades, pesticide usage has shown a sustained growing tendency as a result of the exponential increase of the human population that brings along the need for intensifying food production. In this context, herbicides act as a key component of modern global agricultural systems (Bilal et al., 2019). Diuron [3-(3,4-dichlorophenyl)-1,1-dimethylurea], a phenylurea herbicide, is used extensively to control weeds throughout the world, not only in agriculture but also in urban and industrial scenarios (Liu, 2014). This herbicide is stable to hydrolysis at neutral pH (pH 5–9) and is generally persistent in soil with a half-live of 320–330 days, and thus, can reach water bodies through leaching or surface runoff (Giacomazzi and Cochet, 2004; Langeron et al., 2014; Liu, 2014). Moreover, its main degradation product, 3,4-dichloroaniline (3,4-DCA) exhibits high toxicity and persistence in soils (Giacomazzi and Cochet, 2004; Tasca and Fletcher, 2018). The presence of diuron in the environment may pose a severe threat to human health and aquatic organisms in several ways (Giacomazzi and Cochet, 2004; Huovinen et al., 2015; Liu, 2014; Mansano et al., 2018). To reduce diuron risks and protect water ecosystems, the European Commission has included this herbicide in the priority hazardous substance list (Directive 2013/39/EU) and established environmental quality standards (EQS) in surface waters. Due to its potential effects, diuron is also one of the 92 substances that are currently being assessed by the European Chemicals Agency (ECHA) as a potential endocrine disruptor. Despite worldwide efforts to control its environmental presence, diuron is still frequently detected in surface and ground waters (Kaonga et al., 2015; Rippy et al., 2017; Santos et al., 2015). Thus, scientific efforts are needed to overcome its persistence and recalcitrance (Liu et al., 2018; López-Ramón et al., 2019; Santos et al., 2019). In this setting, the development of techniques to remove diuron and 3,4-DCA from the environment is strongly motivated and urgent.

Bioremediation techniques present two major advantages over both chemical and physical remediation approaches. sustainability and low cost (Azubuike et al., 2016). To date, various microorganisms have been reported to harbor the capacity to simultaneously degrade diuron and 3,4-DCA (Ellegaard-Jensen et al., 2013; Sørensen et al., 2008; Sharma et al., 2010; Villaverde et al., 2017). Most of the studies conducted in this line have focused on bacterial species, while white-rot fungi (WRF), that are metabolically versatile and capable of degrading a wide spectrum of xenobiotics due to their nonspecific lignin-degrading enzymes, have only been marginally investigated (Singh and Singh, 2014). Although different reactors based on WRF have been successfully designed for the degradation of various xenobiotics, their implementation at full-scale still needs to overcome some problems, like bacterial contamination (Mir-Tutusaus et al., 2018). In comparison with other reactors such as stirred tank reactor and fluidized bed reactor, trickle-bed reactors (TBR) offer multiple advantages, such as lower operational costs, greater operational flexibility, and higher sustainability (Luo et al., 2014). Furthermore, the immobilization of *T. versicolor* on wood has been proven to be advantageous to maintain fungal activity and has been successfully used to remove pharmaceuticals in hospital wastewater (Torán et al., 2017).

The purpose of the present study was: i) to screen out one WRF species that could effectively degrade diuron, ii) explore the enzymatic system involved in the degradation, and iii) evaluate the potential transformation products (TPs) formed during the process. A further objective was to estimate the operational variables of a TBR immobilized with the selected fungus for diuron removal, operated under non-sterile conditions, to improve reactor performance and overcome the current limitations for its application in real scenarios.

## 2. Materials and methods

### *2.1. Microorganisms and media*

*Trametes versicolor* ATCC 42530 was acquired from American Type Culture Collection, *Gymnopilus luteofolius* FBCC 466 and *Stropharia rugosoannulata* FBCC 475 were obtained from Fungal Biotechnology Culture Collection (FBCC) of the University of Helsinki (Finland), and *Pleurotus ostreatus* was isolated from a fruiting body collected from rotting wood (Palli et al., 2014) and preserved in our laboratory. Fungal strains were maintained by subculturing every 30 days on 2% (w/v) malt extract plates (pH 4.5) at 25 °C. Blended mycelial suspensions and pellets were prepared using malt extract medium (pH 4.5) as previously described (Blánquez et al., 2004).

The defined medium used in the degradation batch experiments in Erlenmeyer flask consisted of (per liter) 8 g glucose, 3.3 g ammonium tartrate, 1.68 g dimethyl succinate, 10 mL micronutrients, and 100 mL macronutrients (Kirk et al., 1978). pH was adjusted to 4.5.

### *2.2. Chemicals and reagents*

Diuron (purity, ≥ 98%), 3,4-DCA (98%), laccase mediators violuric acid monohydrate (VA, ≥ 97%) and 2,2'-azino-bis (3-ethylbenzothiazoline-6-sulphonic acid) diammonium salt (ABTS, 98%), cytochrome P450 inhibitor 1-aminobenzotriazole (ABT, purity, 98%), and commercial laccase purified from *T. versicolor* (20 AU $mg^{-1}$) were purchased from Sigma-Aldrich (Barcelona, Spain). Diuron-$d_6$, used as a surrogate standard in chemical analyses, was purchased from Toronto Research Chemicals (Ontario, Canada). Hydrated 1-hydroxybenzotriazole (HOBT, ≥ 98%) was obtained from Fluka (Barcelona, Spain). Chromatographic grade acetonitrile was purchased from Carlo Erba Reagents S.A.S (Barcelona, Spain). HPLC-grade methanol, and formic acid (>98%) used as a mobile phase modifier, were obtained from Merck (Darmstadt, Germany). Stock solutions (5 mg $mL^{-1}$) of diuron and 3,4-DCA to be used in *in vivo* and *in vitro* degradation experiments were prepared by appropriate dilution of the substances in ethanol and stored at – 20 °C until use.

### *2.3. Fungal selection for diuron removal*

To select the best fungal species for diuron removal, degradation experiments were performed in 250 mL Erlenmeyer flasks containing 50 mL of defined medium fortified with diuron at a final concentration of 10 mg $L^{-1}$. Briefly, pellets of each fungus were transferred into flasks as inoculum, thereby achieving a concentration of approximately 3.4 g dry weight (DW) $L^{-1}$. Then the cultures were incubated at 25 °C under continuous orbital-shaking (135 rpm) for 7 days. To obviate the influence of photodegradation, the incubation were performed in the dark. Abiotic (uninoculated) and heat-killed culture (121 °C for 30 min) were used as controls. Each set was tested in triplicate. Aliquots were taken at specific time intervals during incubation to measure diuron and glucose concentrations.

### *2.4. Evaluation of the enzymatic system involved in diuron degradation*

*T. versicolor* was selected for subsequent experiments based on its decomposition efficiency. To evaluate the enzymatic system involved in diuron degradation, *in vitro* experiments with *T. versicolor* laccase and *in vivo* experiments with an inhibitor of the cytochrome P450 system were conducted.

Laccase-mediated degradation experiments were performed in 250 mL Erlenmeyer flasks containing 50 mL laccase-sodium malonate dibasic monohydrate solution (250 mM, pH 4.5) at a final enzyme activity of 500 AU $L^{-1}$. Diuron concentration was set to 10 mg $L^{-1}$. Different laccase-mediators including VA, ABTS, and HOBT were added individually to a final concentration of 1 mM (Marco-Urrea et al., 2009) for evaluation of their effects on diuron degradation. Abiotic (no laccase) and flasks without the addition of mediators were also prepared as controls. The flasks were incubated for 3 days on an orbital shaker (135 rpm) at 25 °C in the dark and each experimental condition was conducted in triplicate. At selected times, 1 mL sample aliquot was withdrawn from each flask and mixed with 100 μL of 1 M HCl to stop the reaction. Then, it was filtered through a Millipore Millex-GV PVDF 0.22 μm membrane before the analysis of diuron.

For evaluating the activity of cytochrome P450, the cytochrome P450 inhibitor ABT (dissolved in ethanol) was added to *in vivo* cultures at 5 mM final concentration (Marco-Urrea et al., 2009). Briefly, experiments in triplicate were carried out in 100 mL Erlenmeyer flasks containing 25 mL of defined medium spiked with diuron at a final concentration of 10 mg $L^{-1}$. After inoculating pellets (3.2 g DW $L^{-1}$), cultures were incubated for 7 days at 25 °C in the dark under continuous orbital-shaking (135 rpm). Experimental control was run in parallel in the absence of ABT. Samples were taken at selected times and filtered as mentioned above before the analysis of diuron residues.

### *2.5 Identification of diuron transformation products*

Experiments were conducted in 500 mL Erlenmeyer flasks containing 100 mL of fresh define medium. In brief, after transferring 2.6 g DW $L^{-1}$ of *T. versicolor* pellets, cultures were incubated for 7 days under the same conditions described in section 2.3. Diuron was added at a concentration of 1 mg $L^{-1}$. Abiotic control containing only diuron was also prepared. Each experiment was conducted in triplicate. At selected times, 4 mL of the culture was withdrawn and centrifuged (17, 700 × g, 4 min) at room temperature. Then, 1.5 mL of supernatant was transferred into a 2-mL amber vial containing 75 μL of diuron-$d_6$ solution at a concentration of 10 μg $mL^{-1}$ and kept in the dark at – 20 °C until analysis. Simultaneously, laccase activity was determined during incubation.

Also, another degradation experiment using 3,4-DCA as substrate (10 mg $L^{-1}$) was performed in parallel, to explore the degradation ability of *T. versicolor* towards the main metabolite formed by diuron and elucidate the biochemical pathway of diuron degradation. In these experiments, 3.2 g DW $L^{-1}$ of biomass was employed and all the other conditions were identical to those used in diuron degradation experiments (section 2.3). Samples were taken at selected times and filtered as mentioned above before the analysis of 3,4-DCA residues.

### *2.6. Diuron removal in fungal TBR under non-sterile condition*

As shown in Figure 1, a cylindrical glass TBR (Ø 3 cm, h 25 cm) with a working volume of 160 $cm^3$ was set up. Specifically, 60 g (29 g DW) of autoclaved wet pine wood chips (2 cm × 1 cm × 0.5 cm) were inoculated with blended mycelial suspension and incubated statically for 9 days as described by Torán et al. (2017). Tap water (adjusted to pH 4.5) was fortified with diuron to a final concentration of 10 mg $L^{-1}$, and fed into the reservoir bottle from where it was pumped continuously to the top of the reactor through an external recirculation loop. The hydraulic retention time (HRT) corresponding to the total aqueous volume in the system divided by the influent flow rate ($F_{in}$) was established as 3 days. Humid air was introduced from the top, and the pH was maintained at 4.5 in the reservoir by adding either 1 M HCl or 1 M NaOH, and mixing with a magnetic stirrer.

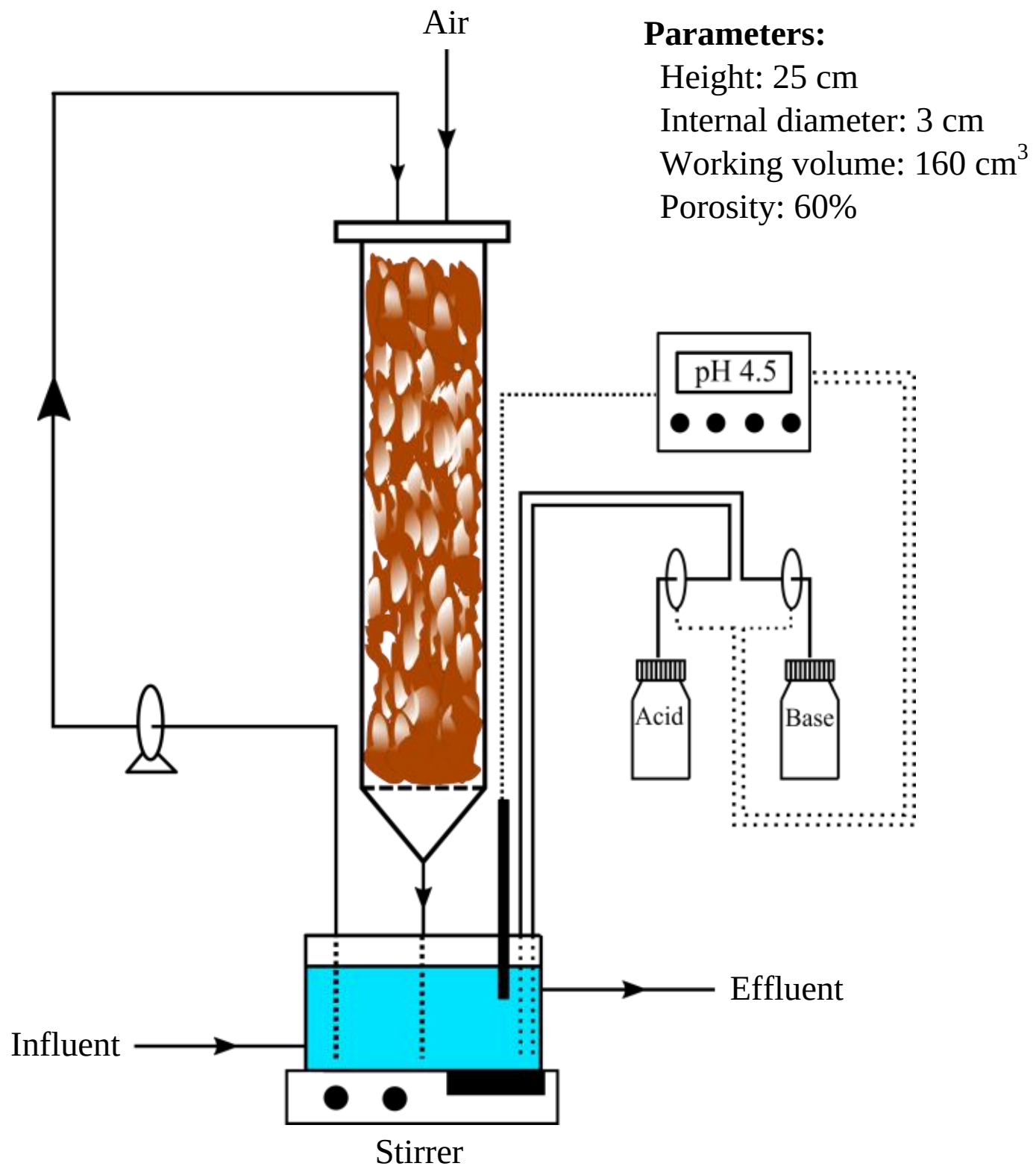

**Figure 1** Schematic representation of the trickle-bed reactor set-up

Response surface methodology (RSM) was applied for assessing the factors that influence diuron removal performance. Given the configuration of this particular system and its variable parameters, recycling ratio (RR), defined as the ratio between recirculating flow rate ($F_r$) and influent flow rate ($F_{in}$), and the ratio of influent flow rate to the working volume of the reactor ($F_{in}/V_R$) were set as independent factors. Correspondingly, the response analyzed was diuron removal at steady state, i.e., after 9 days of continuous treatment. For each experiment, the TBR was newly set up with freshly colonized wood chips. Design Expert V8.06 (Stat-Ease, Inc., MN, USA) was used to design the experiment. A factorial experimental design with 2 factors at 3 levels was employed, as described in Table 1. This translates into 12 experimental runs including 4 replicates.

**Table 1** Factors and levels for response surface methodology

| Factors | Codes | Levels | | |
|---|---|---|---|---|
| | | – 1 | 0 | 1 |
| Recycling ratio (RR) | $X_1$ | 200 | 500 | 800 |
| $F_{in}/V_R$ ($h^{-1}$) | $X_2$ | 0.42 | 1.04 | 1.67 |

## *2.7. Analytical methods*

### *2.7.1. Biomass quantification*

Biomass was determined by the dry weight of pellets. It was obtained through filtration of the culture, followed by washing with sterile distilled water and drying at 105 °C to constant weight.

### *2.7.2. Glucose concentrations*

Samples were filtrated through a nylon membrane (0.45 μm) and then measured with a biochemistry analyzer (2700 select, Yellow Springs Instrument, USA).

### *2.7.3. Laccase activity*

Laccase activity was measured through the oxidation of 2,6-dymethoxyphenol (DMP) by the enzyme as described elsewhere (Wariishi et al., 1992). Activity units per liter (AU $L^{-1}$) are defined as the amount of DMP in μM which is oxidized per minute. The molar extinction coefficient of DMP was 24.8 $mM^{-1}$ $cm^{-1}$.

### *2.7.4. Diuron and 3,4-DCA concentrations*

Diuron and 3,4-DCA residual concentrations were determined using high-performance liquid chromatography (HPLC, Ultimate 3000, Dionex, USA) and UV detection. Briefly, 1 mL of culture-aliquot was withdrawn and filtered (0.22 μm, PVDF) before HPLC analysis. HPLC-UV analysis was performed using a C18 reversed-phase column (Phenomenex®, Kinetex® EVO C18 100 Å, 4.6 mm × 150 mm, 5 μm) at 30 ºC and a mobile phase consisting of acetonitrile and water (40:60, v/v) at a constant flow rate of 0.9 mL $min^{-1}$. Analytes were detected using a wavelength of 252 nm. The sample injection volume was 40 μL.

*2.7.5. Evaluation and identification of TPs*

Analysis of TPs was performed using an ultra-high performance liquid chromatography (UHPLC) system Acquity (Waters, Milford, MA, USA) coupled to a hybrid quadrupole-Orbitrap mass spectrometer Q Exactive (Thermo Fisher Scientific, San Jose, CA, USA), equipped with a heated-electrospray ionization source (HESI). Chromatographic separation was achieved with a Purospher® STAR RP-18 endcapped Hibar® HR (150 × 2.1 mm, 2 μm) column (Merck, Darmstadt, Germany) and a linear gradient of the organic constituent of the mobile phase. The mobile phase employed under positive ionization mode (HESI+) consisted of (A) water and (B) methanol, both containing 0.1% of formic acid (flow rate of 0.2 mL $min^{-1}$), whereas under negative ionization mode (HESI–) a mobile phase of (A) water and (B) acetonitrile (flow rate of 0.3 mL $min^{-1}$) was used. The organic gradient applied was: 0–1 min, 5% B; 3 min, 20% B; 6 min, 80% B; 7 min, 100% B; 9 min, 100% B; 9.5 min, 5% B; 14 min, 5% B. The injection volume was 10 μL.

The specific conditions used in the HESI interface in both polarity modes were: spray voltage, 3.0 kV; sheath gas flow rate, 40 arbitrary units; auxiliary gas, 10 arbitrary units; capillary temperature, 350 ºC, and vaporizer temperature, 400 ºC. Nitrogen (>99.98%) was employed as sheath, auxiliary and sweep gas. Accurate mass detection was conducted in data-dependent acquisition (DDA) mode. First, a full MS scan was done within the *m/z* range 70-1,000 at 70,000 full width at half maximum (FWHM) resolution(at m/z 200). Then, data-dependent MS/MS scan events (17,500 FWHM resolution at m/z 200) were recorded for the five most intense ions (>10e$^5$) detected in each scan, with a normalized collision energy of 40%. Data acquisition was controlled by Xcalibur 2.2 software (Thermo Fisher Scientific).

*2.7.6. HRMS data processing*

The acquired LC-HRMS data were processed using Compound Discover 3.1 software (Thermo Fisher Scientific). Briefly, experimental samples collected at different times were compared, considering samples collected at t=0 as the control. *m/z* features in the different samples were aligned and deconvoluted using 2 min as maximum retention time shift and 5 ppm of mass tolerance. Then, they were grouped and assigned to a specific molecular ion, for which an elemental composition was predicted. In parallel, a search by formula or mass was performed in various MS libraries and compound databases (ChemSpider, mzCloud, mzVault) for the assignment of a potential compound identity. This workflow produced a list of peaksthat was manually revised to identify real TPs, i.e., those peaks that were only present in samples collected after 2 and/or 7 days of degradation, and absent at time 0. Once identified, the molecular structures proposed by the software were evaluated according to the elemental composition of the molecular and fragment ions, fragment rationalization (assisted by fragment ion search scoring), and isotopic patterns.

#### *2.7.7. Data analysis*

The mean and standard deviation (SD) of data were calculated and subjected to analysis of variance (ANOVA). Statistical significance was determined using SPSS V22.0.

## 3. Results and discussion

### *3.1. Degradation of diuron by different white-rot fungi*

Firstly, the ability of different white-rot fungi to degrade diuron was evaluated. As shown in Figure 2, diuron in abiotic controls showed high chemical stability after 7 days of incubation. This means that any removal observed in the experimental treatment must be exclusively attributed to fungal sorption and biodegradation mechanisms. The four tested microorganisms demonstrated different diuron elimination efficiencies, ranging from 18% to 83%. Although there were small differences, the variation patterns of diuron concentration in the different heat-killed fungal cultures were essentially similar, showing an initial fast drop due to sorption onto the biomass, after which concentration remains stable. Thus, sorption contributed to diuron elimination to some extent. This is in agreement with data reported by Lucas et al. (2018) in similar WRF elimination experiments conducted with pharmaceuticals. On average, diuron adsorption observed in this study (16%) was more than double than that observed for other xenobiotics with similar or higher hydrophobicity than diuron (Log Kow 2.68), i.e., venlafaxine (Log Kow 3.28) and carbamazepine (2.45) (Lucas et al., 2018). Thus, additional physical-chemical characteristics may affect the adsorption of the compound onto the biomass (e.g, pKa, or water solubility).

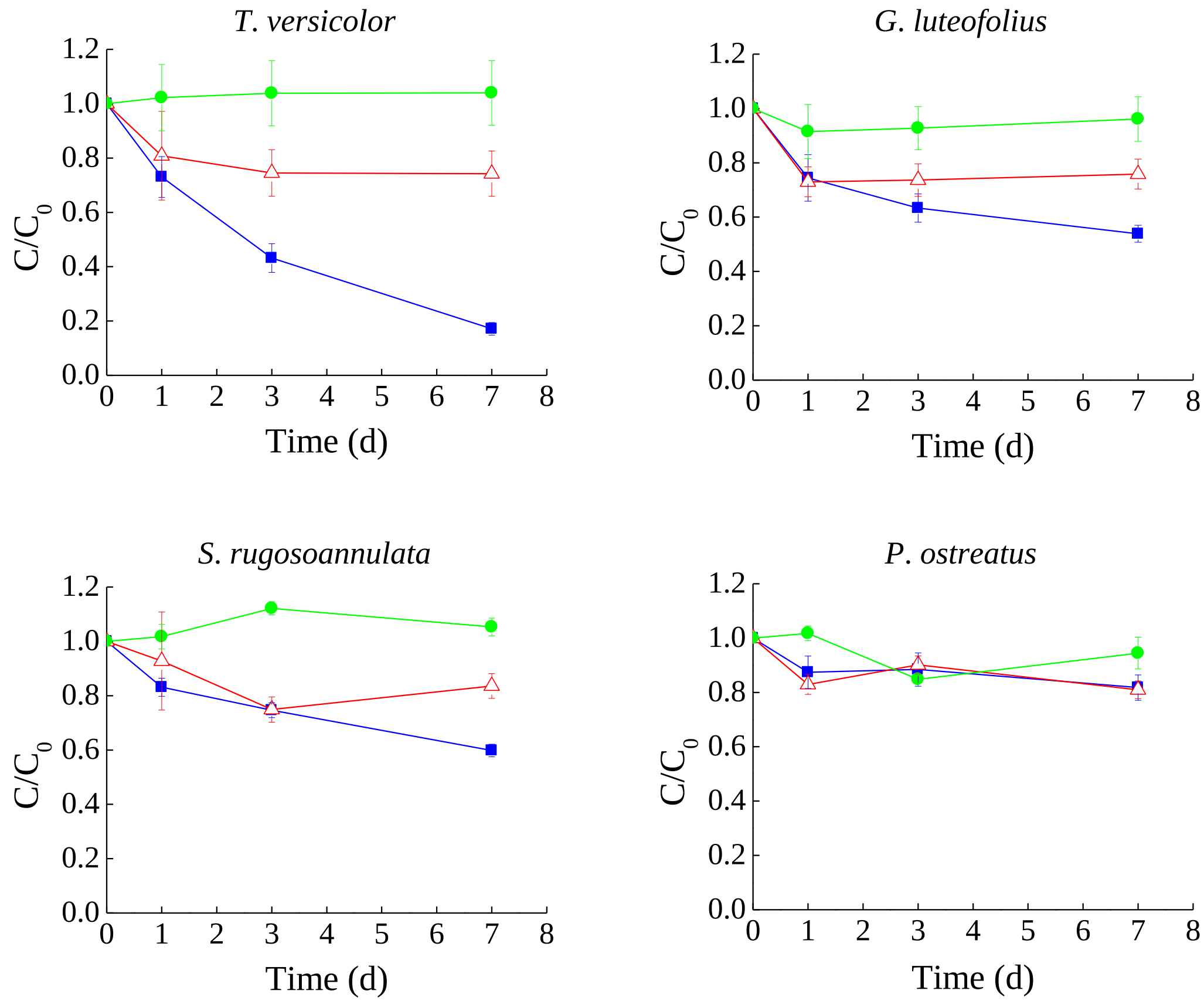

**Figure 2** Time-course degradation of diuron by different fungi. C represents the residual concentration of diuron in the sample (mg $L^{-1}$), and $C_0$ corresponds to the initial concentration of diuron in the sample (mg $L^{-1}$); *Blue lines with filled squares*, experimental; *red lines with empty triangles*, killed control; *green lines with filled circles*, abiotic. Average values of three replicates with the corresponding standard deviation are shown.

As for glucose, its concentration plunged to almost zero after 7 days of incubation in *T. versicolor*, *G. luteofolius,* and *S. rugosoannulata* cultures (Table 2). Although the capability of *P. ostreatus* in degrading different xenobiotics such as plastic, organochlorine pesticides, and polycyclic aromatic hydrocarbons has been well documented (Bhattacharya et al., 2014; da Luz et al., 2013; Purnomo et al., 2010; Purnomo et al., 2017), an exceptional scenario was found in diuron degradation, as it presented a much lower consumption of glucose than the other investigated fungi, with 4.85 g $L^{-1}$ glucose remaining in solution at the end of the experiment. This is indicative of a lower metabolism of this fungal strain as compared to the others, which results in nearly no biodegradation (as shown in Figure 2). A similar observation was reported regarding the degradation of venlafaxine by this species in an analogous medium (Llorca et al., 2019), in spite that it showed excellent performance for the removal of other pharmaceuticals, namely, diclofenac, ketoprofen, and atenolol, in sterile hospital wastewater (Palli et al., 2017). Hence, further research is needed to address the mechanisms behind this selective performance.

**Table 2** Time-course of glucose concentrations in the experimental culture media (g $L^{-1}$)

| Fungus | Incubation time (d) | | | |
|---|---|---|---|---|
| | 0 | 1 | 3 | 7 |
| *T. versicolor* | 7.00 ± 0.13 | 5.35 ± 0.16 | 1.83 ± 0.40 | < 0.01 |
| *G. luteofolius* | 6.63 ± 0.10 | 4.77 ± 0.02 | 0.54 ± 0.29 | < 0.01 |
| *S. rugosoannulata* | 7.02 ± 0.13 | 4.34 ± 0.02 | < 0.01 | ND |
| *P. ostreatus* | 7.01 ± 0.12 | 6.21 ± 0.09 | 5.67 ± 0.05 | 4.85 ± 0.10 |

Note: Each value of glucose concentration represents the mean of triplicate measurements ± SD. ND: no detected

This is the first study that confirms the potential of *S. rugosoannulata* to degrade diuron, thus enriching the diuron-degrading microbe pool. Considering the better performance in diuron degradation by *T. versicolor* compared to the other investigated WRF species, it was selected for further research.

### *3.2. Role of laccase and cytochrome P450 inhibitor in the degradation of diuron*

Several *in vitro* experiments were carried out to investigate whether laccase and laccase-mediator systems are involved in diuron degradation by *T. versicolor*. Laccase activity was found to reach a maximum level of 27.68 AU $L^{-1}$ during diuron degradation (TP formation) experiments. Moreover, results of *in vitro* degradation experiments showed that diuron concentrations remained constant throughout the incubation period and were similar to those observed in the abiotic control (Table 3), despite substantial laccase activity (500 AU $L^{-1}$) was introduced. These results indicate that laccase is not involved in the first step of diuron degradation. Nonetheless, an opposite conclusion was drawn in a previous study (Coelho-Moreira et al., 2018), in which diuron was successfully depleted in the presence of crude laccase from *G. lucidum,* and improved yields were obtained after adding synthetic mediators, namely, ABTS and acetylacetone. Furthermore, Coelho-Moreira et al. (2018) reported that diuron acted as a laccase inducer. The genetic difference among genera and species of WRF could explain the discrepancy of the results obtained. Likewise, the effect of different redox potentials between the laccase enzyme and the monophenolic substrate may be a key factor (Glazunova et al., 2018). In this regard, it is also worthy to point out that low levels of Mn peroxidase were detected in culture filtrate from *G. lucidum* (Coelho-Moreira et al., 2018).

**Table 3** Effect of laccase on diuron degradation in the presence and absence of mediators

| Time (h) | Diuron (mg $L^{-1}$) | | | | |
|---|---|---|---|---|---|
| | Abiotic control | Laccase | VA | ABTS | HOBT |
| 0 | 10.01 ± 0.41 | 9.69 ± 0.91 | 9.02 ± 0.31 | 9.58 ± 0.72 | 9.41 ± 0.29 |
| 8 | 9.83 ± 0.38 | 9.74 ± 0.59 | 8.96 ± 0.47 | 10.35 ± 0.21 | 9.25 ± 0.71 |
| 24 | 9.45 ± 0.75 | 9.40 ± 0.60 | 9.69 ± 0.49 | 9.66 ± 0.35 | 10.60 ± 0.83 |
| 72 | 9.32 ± 1.10 | 9.01 ± 0.89 | 9.54 ± 1.10 | 10.74 ± 0.16 | 9.13 ± 1.47 |

Note: Each value of laccase concentration represents the mean of triplicate measurements ± SD. VA, violuric acid monohydrate; ABTS, 2,2'-azino-bis (3-ethylbenzothiazoline-6-sulphonic acid) diammonium salt; HOBT, hydrated 1-hydroxybenzotriazole

The activity of the cytochrome P450 enzymatic system in diuron degradation was evaluated *in vivo* experiments after the addition of the cytochrome P450 inhibitor 1-aminobenzotriazole. The presence of the cytochrome P450 inhibitor in the culture hindered diuron degradation as compared to the inhibitor-free culture system, although some removal occurred initially, which was probably attributed to sorption processes (Figure 3). This finding is in agreement with previous reports that showed that the addition of 1-aminobenzotriazole significantly inhibited diuron degradation by *P. chrysosporium* and *G. lucidum*, as well as the metabolites produced in the process (Coelho-Moreira et al., 2013, 2018).

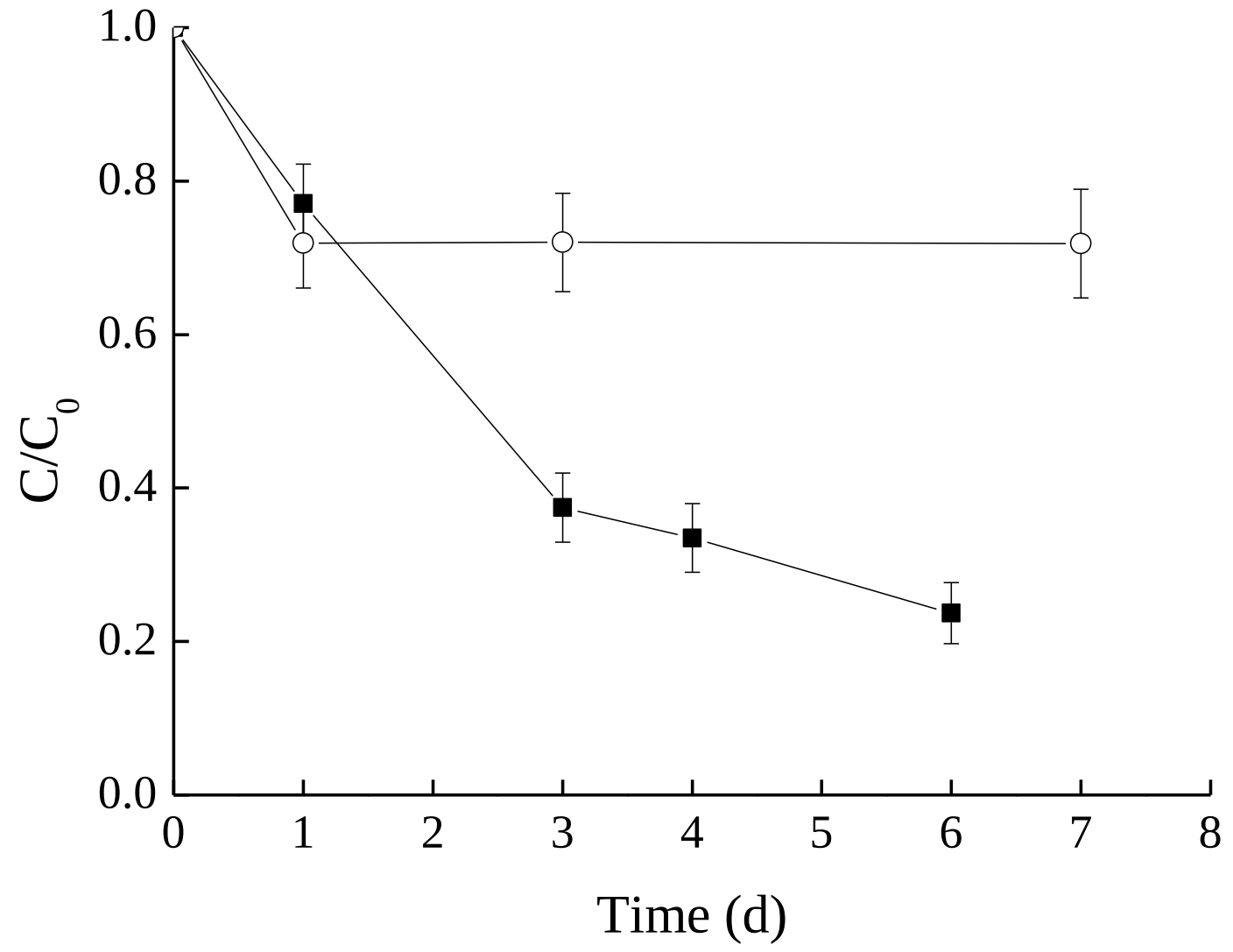

**Figure 3** Effect of the cytochrome P450 inhibitor 1-aminobenzotriazole on the degradation of diuron by *T. versicolor*. C represents the residual concentration of diuron in the sample (mg $L^{-1}$), and $C_0$ corresponds to the initial concentration of diuron in the sample (mg $L^{-1}$); *Filled squares*, inhibitor-free; *empty circles*, inhibitor. Average values of three replicates with the corresponding standard deviation are shown.

### *3.3. TPs generated during degradation of diuron by T. versicolor*

Main TPs generated during the degradation of diuron by *T. versicolor* and detected using LC-HRMS are shown in Table 4. In total five ions were identified as TPs; however, logical tentative structures could be only proposed for three of them with an identification confidence level of 3 (CL3) according to Schymanski's scale (Schymanski et al., 2014). The remaining two TPs were appointed as "unequivocal molecular formulae" (identification confidence level of 4), as no sufficient evidence existed to propose possible structures. The CL3 TPs were TP248 3-(3,4-dichlorophenyl)-1-hydroxymethyl-1-methylurea (DCPHMU), TP218 3-(3,4-dichlorophenyl)-1-methylurea (DCPMU), and TP204 3,4-dichlorophenylurea (DCPU).

TP218 (DCPMU) and TP204 (DCPU) have been previously reported as aerobic degradation by-products of diuron regardless of the organism (bacteria or fungi) used for its degradation. They are formed after successive N-demethylations reactions of diuron (Badawi et al., 2009; Coelho-Moreira et al., 2013, 2018; Ellegaard-Jensen et al., 2014; Sørensen et al., 2008). Demethylation of diuron may occur intracellularly, as diuron and these demethylated metabolites were found to be present in mycelial extracts of *G. lucidum* (Coelho-Moreira et al., 2018). This is also in agreement with the results obtained in the laccase experiments conducted in this study. All identified TPs remained in solution after 7 days of treatment; however, TP204 (DCPU), TP137 and TP195 were the ones that presented an increasing trend by the end of the experiment. Thus, TP248 (DCPHMU) and TP218 (DCPMU) can be considered as intermediate byproducts that further degrade in contact with the fungus (Figure 4). TP248 is believed to be formed after carbon hydroxylation at the tertiary amine moiety. This reaction may be mediated by the cytochrome P450 system, according to the enzymatic exploration results and to a previous study on the detoxification of chlortoluron by the grass weed *A. myosuroides* (Hall et al., 1995). This is the first report of the formation of TP248 during microbial degradation of diuron.

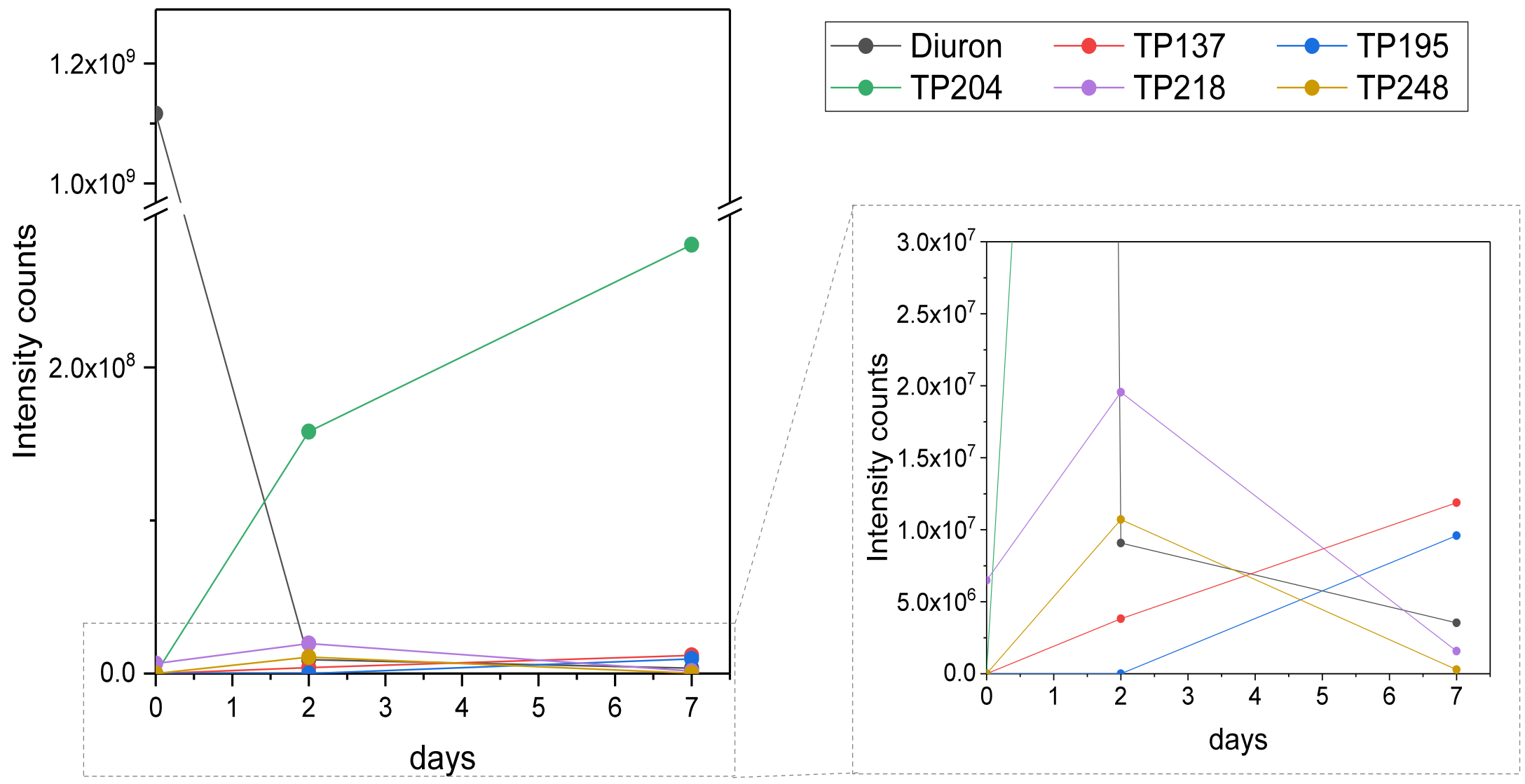

**Figure 4.** Evolution of the TPs identified during *T. versicolor*-mediated degradation of diuron

**Table 4** Transformation products formed during diuron degradation by *T. versicolor*

| CODE | $t_R$ (min) | HESI mode | Full Scan *m/z* | Full Scan Formula | Full Scan RDB | Full Scan Δm (ppm) | MS/MS m/z | MS/MS Formula | MS/MS RDB | MS/MS Δ (ppm) | Suspect identity (Confidence level) | Chemical structure |
|---|---|---|---|---|---|---|---|---|---|---|---|---|
| TP248 | 6.7 | – | 247.0030 | $C_9H_9O_2N_2Cl_2$ | 5.5 | -2.1 | 159.9708 | $C_6H_4NCl_2$ | 4.5 | – 4.6 | DCPHMU (CL3) | 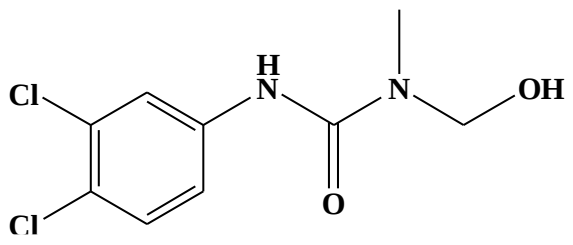 |
| TP218 | 8.6 | + | 219.0093 | $C_8H_9Cl_2ON_2$ | 4.5 | 3.3 | 161.9876 | $C_6H_6Cl_2N$ | 3.5 | 3.8 | DCPMU (CL3) | 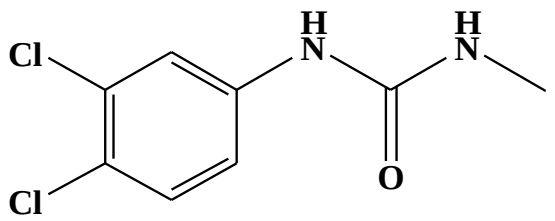 |
| TP204 | 8.5 | + | 204.9937 | $C_7H_6Cl_2N_2O$ | 4.5 | 3.5 | 161.9877 | $C_6H_6Cl_2N$ | 3.5 | 3.4 | DCPU (CL3) | 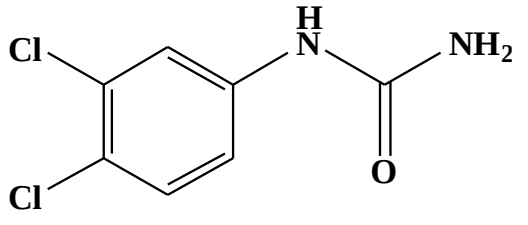 |
| | | | | | | | 159.9728 | $C_6H_4Cl_2N$ | 4.5 | 8.2 | | |
| | | | | | | | 92.0503 | $C_6H_6N$ | 4.5 | 8.5 | | |
| TP137 | 2.9 | + | 138.0556 | $C_7H_8O_2N$ | 4.5 | 4.4 | 110.0608 | $C_6H_8ON$ | 3.5 | 7.1 | n/a (CL4) | |
| TP195 | 4.3 | + | 196.0612 | $C_9H_{10}O_4N$ | 5.5 | 4.0 | 178.0508 | $C_9H_8O_3N$ | 6.5 | 3.6 | n/a (CL4) | |
| | | | | | | | 150.0556 | $C_8H_8O_2N$ | 5.5 | 4.5 | | |

$t_R$: chromatographic retention time, HESI, heated-electrospray ionization; Δm, mass measurement error; RDB, ring and double-bound equivalents 3,4-DCA has been widely reported to be a diuron TP that could be generated after the amide bond hydrolysis of DCPU; however, it was not detected in the experimental culture media in this study. Since 3,4-DCA is much more toxic than diuron, the ability of *T. versicolor* to remove this TP was also evaluated. A degradation experiment using 3,4-DCA as substrate was conducted. *T. versicolor* showed an exceptional ability to degrade 3,4-DCA Initial 3,4-DCA concentrations (10 mg $L^{-1}$) were degraded in less than 24 h (Figure 5). This may explain why 3,4-DCA was not identified as a diuron TP in this study. The biodegradation of 3,4-DCA under aerobic conditions has been extensively described, and it could be reduced to a dechlorination step followed by ring-cleavage (Giacomazzi and Cochet, 2004; Tasca and Fletcher, 2018).

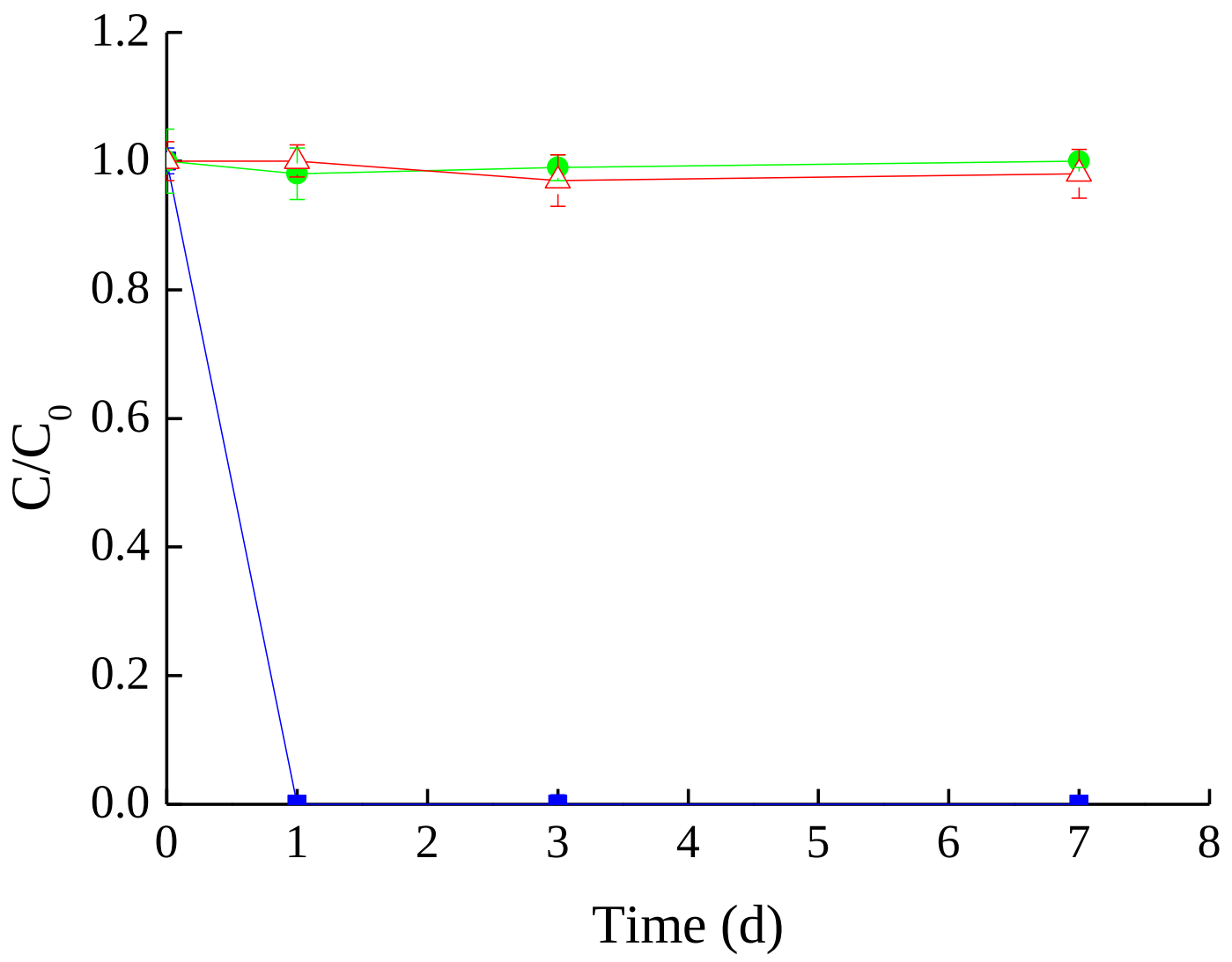

**Figure 5** Time-course degradation of 3,4-DCA by *T. versicolor*. C represents the residual concentration of 3,4-DCA in the sample (mg $L^{-1}$), and $C_0$ corresponds to the initial concentration of 3,4-DCA in the sample (mg $L^{-1}$); *Blue lines with filled squares*, experimental; *red lines with empty triangles*, killed control; *green lines with filled circles*, abiotic. Average values of three replicates with the corresponding standard deviation are shown.

Based on our results and previous findings, a pathway of diuron degradation by *T. versicolor* is proposed in Figure 6; however, further investigation is still needed to understand the entire metabolic pathway.

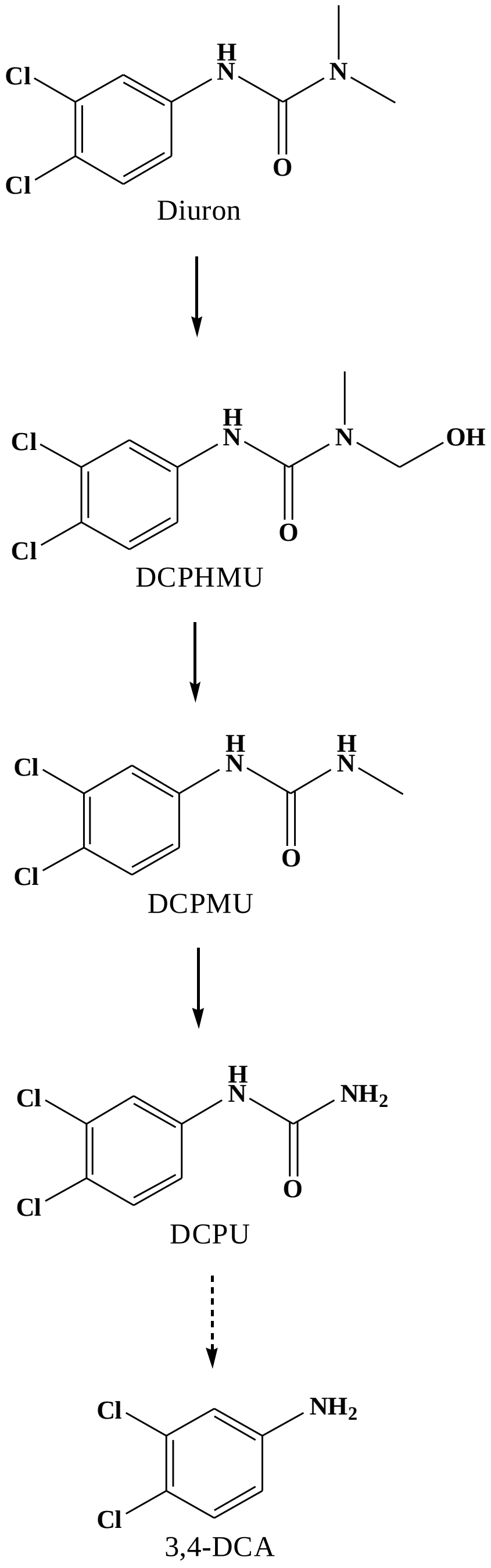

**Figure 6** Proposed pathway of diuron degradation by *T. versicolor*. *Full lines*, the reactions observed in this study; *dashed lines*, the reaction described in the literature

*3.4. Influence of operational variables on diuron removal using TBR under non-sterile condition*

TBR performance is difficult to describe using a mathematical model due to the various physical and biological processes involved in it. On one side, the contribution of the wood is not limited to the immobilization of the biomass. It also plays an important role in pollutant removal due to adsorption, which occurs mainly in the early stages of a continuous treatment process. On the other side, wood is a nutrient source that promotes fungal growth. The water is distributed from the bottom to the top of the TBR and flows through a bed with a remarkably high porosity (60%), which is mainly occupied by air. Thus, the removal yield will depend mainly on the contact time between the immobilized biomass and the liquid phase. The biomass is directly related to the mass of the wood, referred to as the working volume of the bioreactor ($V_R$) (Fig. 1). Since the working volume ($V_R$) and the HRT were fixed, the factors amenable to modification were narrowed down to the recirculation flow rate ($F_r$), the influent flow rate ($F_{in}$), and the water volume in the reservoir tank ($V_r$). The number of times that the water is recirculated through the bioreactor is designated as RR ($F_r/F_{in}$). The HRT, defined as the ratio between the volume of water in the system (i.e., the volume of water in the reservoir tank ($V_r$)) and the $F_{in}$, was fixed to 3 days according to previous experiences (Torán et al., 2017). The ratio between $F_{in}$ and $V_R$ stands for the pollutant load to the system. Therefore, the factors RR ($X_1$) and $F_{in}/V_R$ ($X_2$) were selected for studying each effect, as well as their interaction on the removal of diuron. According to the factorial experimental design with 3 levels for each of these two factors, 12 trials were performed using tap water fortified with diuron, and each experimental result along with model predicted values are given in Table 5.

**Table 5** Central Composite Design matrix showing actual and predicted values for diuron removal

| Run | Factors in coded value | | Response (Diuron removal, %) | |
|---|---|---|---|---|
| | $X_1$ | $X_2$ | Actual | Predicted |
| 1 | 0 | – 1 | 74.4 | 74.1 |
| 2 | 1 | – 1 | 67.3 | 66.9 |
| 3 | 1 | 0 | 44.7 | 44.1 |
| 4 | 0 | 0 | 56.7 | 56.1 |
| 5 | – 1 | 0 | 64.6 | 68.2 |
| 6 | – 1 | – 1 | 83.0 | 81.2 |
| 7 | 1 | 1 | 26.5 | 21.3 |
| 8 | – 1 | 1 | 62.6 | 55.2 |
| 9 | 0 | 0 | 29.5 | 38.2 |
| 10 | – 1 | – 1 | 84.0 | 81.2 |
| 11 | – 1 | – 1 | 82.0 | 81.2 |
| 12 | – 1 | – 1 | 78.0 | 81.2 |

In each trial, the packing material was renovated to avoid fungus aging and adverse effect of saturated wood on adsorption. The response of each run was the mean removal value obtained during the steady state, which was achieved from day 9 to day 18 of continuous operation. Based on the fitness of the data obtained, the following linear regression model for diuron removal in the TBR system is proposed:

$$Y = 56.61 - 11.98X_1 - 17.72X_2 - 5.05X_1X_2 \quad (1)$$

To further verify the goodness of the proposed model for diuron removal, analysis of variance (ANOVA) was applied (Table 6). The results obtained indicate that the proposed model is significant with a confidence level of 95% ($F = 58.54$, $p < 0.0001$), which also addressed its stability. The adjusted $R^2$ was 0.9401, indicating that 94 % of the data variance could be explained by the studied factors (with a level of confidence of 95%). The two factors and the interaction between them were also significant to predict diuron removal ($p < 0.05$) on which the predictor $F_{in}/V_R$ demonstrate highest effect. The lack of fit was non-significant, which evidences the good correlation between the experimental factors or predictors and the response variable. Besides, Table 5 shows the agreement between the experimental data and those predicted by the model. Figure 7 illustrates that the decrease in RR and $F_{in}/V_R$ enhanced the removal of diuron. Therefore, it is clear to infer here the improvement of yield that can be obtained by setting those two factors at the lowest possible levels, having in mind the following condition as $V_R$ is constant.

**Table 6** Analysis of variance of the linear regression model proposed

| Source | Sum of squares | Degrees of freedom | Mean square | F | *p* |
|---|---|---|---|---|---|
| Model | 4, 214.66 | 3 | 1, 404.89 | 58.54 | < 0.0001* |
| $X_1$ | 1, 028.52 | 1 | 1, 028.52 | 42.86 | 0.0002* |
| $X_2$ | 2, 248.02 | 1 | 2, 248.02 | 93.68 | < 0.0001* |
| $X_1X_2$ | 134.73 | 1 | 134.73 | 5.61 | 0.0454* |
| Residual | 191.98 | 8 | 24.00 | | |
| Lack of fit | 171.23 | 5 | 34.25 | 4.95 | 0.1091 |
| Pure error | 20.75 | 3 | 6.92 | | |
| Total | 4, 406.64 | 11 | | | |

*Significant ($p < 0.05$)

$R^2 = 0.9564$

$R^2_{adj} = 0.9401$

$$\frac{\Delta F_r}{F_r} < \frac{\Delta F_{in}}{F_{in}} \quad (2)$$

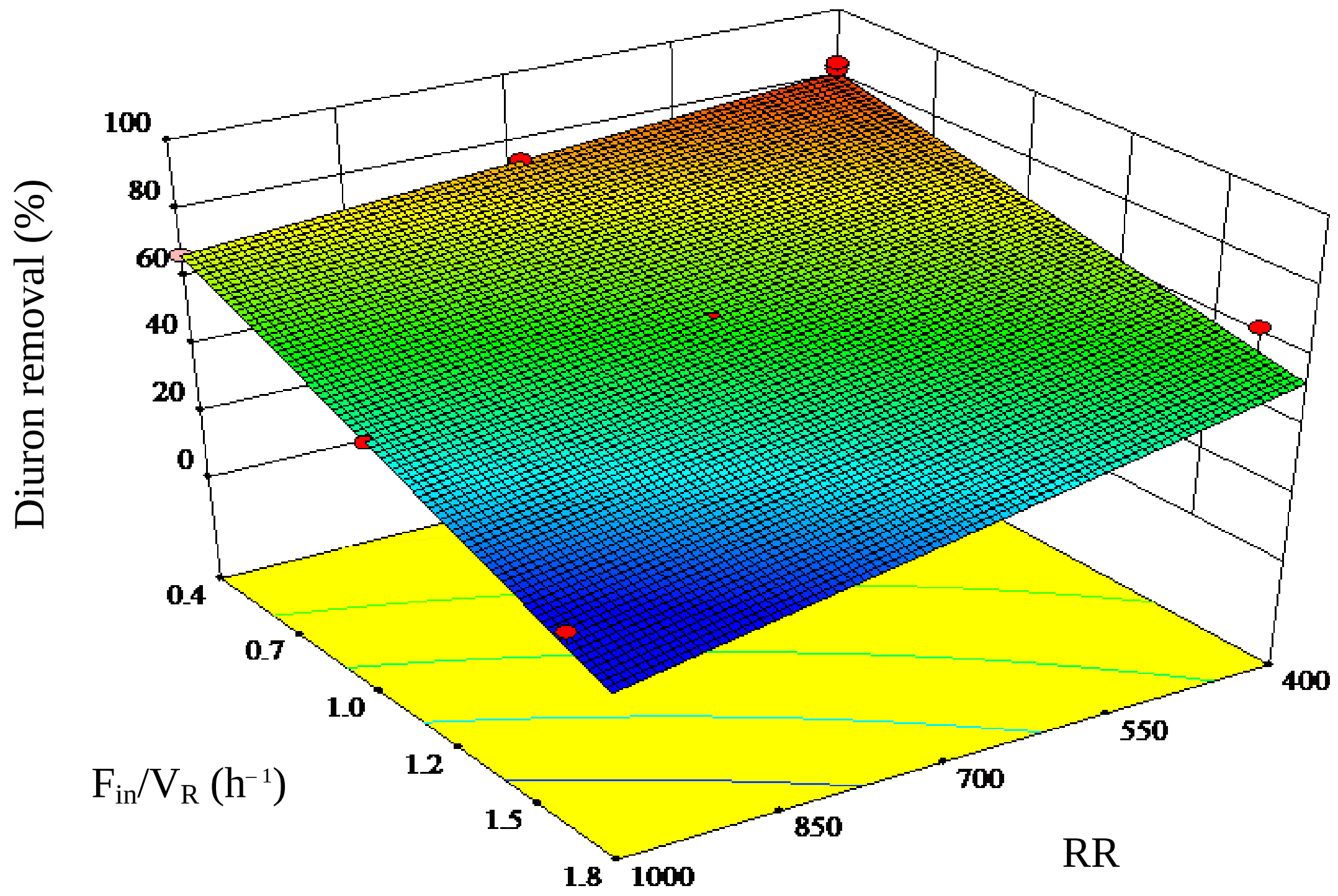

**Figure 7** Response surface showing the effects of RR and $F_{in}/V_R$ on diuron removal

The results evidenced, as expected, that the lower the volumetric load ($F_{in}/V_R$) the higher the pollutant removal, but also that lower recirculating rates entail better removal yields. The RR is related to the times that the water is passed through the bioreactor. In this regard, according to our results, and contrary to expected, low RR values result in better performance than high RR values. This could be explained by the fact that the highest contact time between the immobilized fungus and the pollutant may contribute considerably to increase the removal efficiency, which is also consistent with the fact that the cytochrome P450 system was the one found to be involved in diuron degradation instead of extracellular enzymes. High RR requires high recirculating flow rates promoting a coalescing effect between water drops, which increases the crossing velocity through the column and consequently results in a reduction of the contact time with the fungus. However, a lower value of diuron removal (72% vs 81.2%) was obtained in an additional experiment using RR of 200 and optimal $F_{in}/V_R$. This result indicates that better removal could be achieved by enlarging the experimental domain to the corner where the optimum is. In summary, to design an effective TBR is important to ensure a high contact time between the water and the fungus. This will depend on the operational parameters that further influence the water crossing velocity through the bioreactor.

## 4. Conclusions

Of the four fungi species investigated, *T. versicolor* was found to be the one that most efficiently degraded both diuron and its main byproduct 3,4-DCA. The cytochrome P450 enzymatic system was found to be the one catalyzing diuron removal. Up to five TPs were formed during the process, and structures could be tentatively proposed for three of them. One intermediate TP, namely DCPHMU, was detected for the first time, providing a better understanding of the occurring demethylation process. The effect of operational variables on diuron removal from water using a trickle-bed reactor was also evaluated. The contact time between the immobilized fungus and the water played an important role in diuron removal. This represents a pioneering step forward to overcome key barriers to upscale this type of bioreactor and apply this technology to the bioremediation for real diuron-contaminated waters since the TBR tested in this study was operated under non-sterile conditions.

**Acknowledgment**

This work has been supported by the Spanish Ministry of Economy and Competitiveness State Research Agency (CTM2016-75587-C2-1-R and CTM2016-75587-C2-2-R) and co-financed by the European Union through the European Regional Development Fund (ERDF) and the Horizon 2020 research and innovation WATERPROTECT project (727450). This work was partly supported by the Generalitat de Catalunya (Consolidate Research Group 2017-SGR-14) and the Ministry of Science and Innovation (Project CEX2018-000794-S). The Department of Chemical, Biological and Environmental Engineering of the Universitat Autònoma de Barcelona is a member of the Xarxa de Referència en Biotecnologia de la Generalitat de Catalunya. K. Hu acknowledges the financial support from the Chinese Scholarship Council.

**Conflict of interest**

We declare that no conflict of interest exists in the submission of this manuscript.